# Miniature Plasma Source for In-Situ Scanner Cleaning

Mark van de Kerkhof[a,b], Edgar Osorio[b], Vladimir Krivtsun[c], Maxim Spiridonov[c], Viacheslav Medvedev[c], Dmitry Astakhov[c]

[a] Department of Applied Physics, Eindhoven University of Technology, Eindhoven, The Netherlands
[b] ASML, Veldhoven, The Netherlands
[c] Institute for Spectroscopy of the Russian Academy of Sciences, Troitsk, Moscow, Russia

**ABSTRACT**

EUV Lithography is the technology of choice for High-Volume Manufacturing (HVM) of sub-10nm lithography. One of the challenges is to enable in-situ cleaning of functional surfaces such as sensors, fiducials and interferometer mirrors without opening the scanner tool.
Thermally created hydrogen radicals have been successfully used for this purpose. These sources have a limited cleaning speed and relatively high thermal load to the surface being cleaned. Here we present an alternative plasma-based technique to simultaneously create hydrogen radicals and hydrogen ions. This results in significantly improved cleaning speed while simultaneously reducing the overall thermal load.
As an additional benefit, this plasma source has a minimized and flexible building volume to allow easy integration into various locations in the EUV lithographic scanner.

**Keywords**: EUV Lithography, EUV Scanner, Inductively Coupled Plasma, Plasma Cleaning

I. INTRODUCTION

EUV Lithography, using a wavelength of 13.5 nm, has established itself as the technology of choice for High-Volume Manufacturing (HVM) of 5 nm node and beyond, ensuring that shrinking of semiconductor devices (known as Moore's law) will continue for the coming years [1]. Even with the outstanding imaging and overlay capability of the current EUV scanners [1], device output and yield can still be affected adversely by other factors, such as molecular or particulate contamination on critical imaging surfaces [2]. The EUV scanner operates in near-vacuum (~1-10 Pa) but may have relatively high partial pressures of hydrocarbons around the wafer stage, as a result of outgassing of organic resist components [3]. As a consequence, integrated sensors and mirrors may get contaminated, which may result in drifts in measured dose, position or focus. Currently hydrogen radical generators (HRG) are used for cleaning this carbon-containing contamination when needed. The HRG principle works by thermally generating hydrogen radicals through a hot tungsten filament [4], whereupon the hydrogen radicals will react with carbon to form volatile methane [5]. Drawbacks of the HRG are the relatively low etching rate of hydrogen radicals, of about 1 nm/min, and heating of the surface to be cleaned by radical recombination, of about 1 kW/m$^2$, which requires subsequent cooling in order of minutes in a near-vacuum scanner system before the system is in a stable thermal situation again.

It is well-known from fusion research that the etch yield of hydrogen can be increased by several orders of magnitude by combining radicals with low-energy ions [6]. This synergetic effect is called chemical sputtering [7]. The higher etch yield and higher efficiency result in higher cleaning rate and consequently in lower cleaning time, and thereby lower total heat load per nm cleaned.

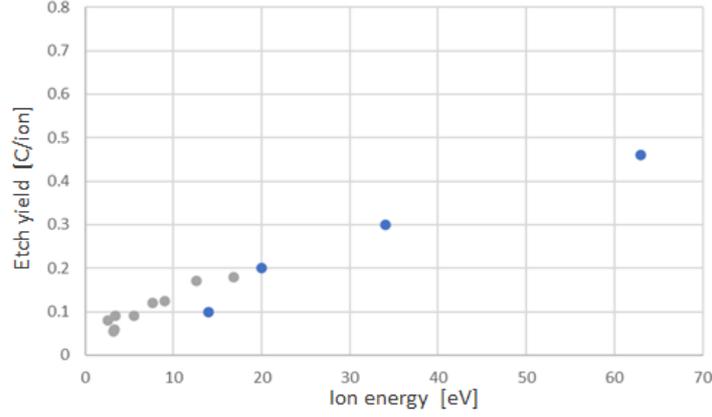

**Figure 1: Carbon etch yield as function of average ion energy in off-line plasma. Blue dots: Surface Wave Discharge at 3 Pa – data courtesy of A. Dolgov [8]. Grey dots: additional data points from internal offline setup (Sairem Aurawave ECR, 20-100W in 5 Pa $H_2$) [9].**

The total cleaning rate $CR$ is a combination of chemical sputtering by plasma and the contribution of radical etching, but may be approximated by the chemical sputtering term, since the chemical sputtering yield is several orders of magnitude higher than the radical etch yield [10]:

$$CR = \frac{M_C}{N_A \rho_C}\left(\underbrace{Y(H^+|H)\varphi_{ion} + Y_{rad}\varphi_{rad}}_{plasma}\;\underbrace{}_{radicals}\right) \cong \frac{M_C}{N_A \rho_C} \cdot Y(H^+|H) \cdot \varphi_{ion} \qquad (1)$$

where $M_C$ is the atomic mass of carbon (12 g/mol), $N_A$ is Avogadro's constant, $\rho_c$ is density of carbon ($2 \cdot 10^3$ kg/m³), $\varphi_{ion}$ is the ion flux, $\varphi_{rad}$ is the radical flux. The etch yield $Y(H^+|H)$ is in order of 0.1-0.5 C/ion for 10-60 eV ions, and has no clear threshold at lower energies, as shown in Figure 1; as the yield scales linearly with the ion energy, the ion energy distribution may be approximated by the average ion energy.

The heat load of the hydrogen plasma towards the surface consists of two contributions, from ions and from radicals. The energy released when an ion hits the surface consists of two parts. Firstly, the ions strike the surface with a kinetic energy $E_{ion}$ in order of 10 eV, which is dissipated in the collision. Secondly, the ion recombines with an electron, releasing the recombination energy $E_{rec,i}$ which for $H_3^+$ is roughly 12-14 eV (depending on the vibrational excitation state of the resulting $H_2$ molecule [11]). When two radicals recombine at the surface to form $H_2$, a recombination energy $E_{rec,H}$ is released, of 2.3 eV per radical (or 4.5 eV total per $H_2$). Hence, the total heat load of the plasma cleaning is:

$$H_{plasma} = \underbrace{(E_{ion} + E_{rec,i})\varphi_{ion}}_{ions} + \underbrace{E_{rec,H}\varphi_{rad}}_{radicals} = (E_{ion} + E_{rec,i} + RE_{rec,H})\varphi_{ion}. \qquad (2)$$

With $R$ the flux ratio of radicals to ions. Assuming typical values for the plasma of $\varphi_{ion} = 1 \cdot 10^{20}$ ions/m²s, $E_{ion}$ = 10 eV, and $R$ = 5, the cleaning rate may be estimated from equation 1 at 7 nm/min, while the heat load may be estimated from equation 2 to be 0.5 kW/m². It should be noted that in general ion energies should not exceed values in order of 20 eV to avoid sputtering damage to the substrate to be cleaned [12]. Table 1 shows a comparison between plasma and HRG.

**Table 1: Summary comparison of cleaning and heating by HRG and hydrogen plasma.**

|  | Cleaning rate (nm/min) | Time to clean 1 nm carbon (s) | Heat load (kW/m2) | Cumulative heat load (kJ/m2) |
|---|---|---|---|---|
| Hydrogen Radical Generator (HRG) | 1 | 60 | 1 | 60 |
| Hydrogen plasma | 2-10 | 6-30 | 0.1-0.5 | ~3 |

Several types of plasma technologies may be considered, but the combination of requirements of small form factor and relatively high ion flux at low ion energies resulted in a choice to investigate the possibility to use an inductively coupled plasma source (ICP). ICP sources use a magnetic field to couple energy into a gas to create and sustain a plasma. The basic driving component of an ICP is a coil, through which a MHz RF current is supplied. The oscillating current generates an oscillating electromagnetic field within and around the coil. Electrons pick up energy from the oscillating electric field, leading to ionization of $H_2$ gas molecules. This is schematically depicted in Figure 2.

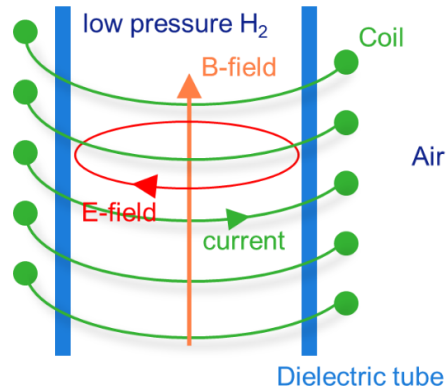

**Figure 2: Schematic overview of working principle of ICP plasma.**

Other configurations of ICP exist. For instance, in a planar arrangement a planar coil is placed against a dielectric window to excite the gas on the other side of the window. A general benefit of ICP is a relatively large ion flux at a relatively low ion energy in order of 10 eV.

## II. MINIATURE ICP DESIGN AND MANUFACTURING

The current mini-ICP design was inspired by the earlier work of Hopwood and Yin [13]. It is based on a planar spiral coil mounted against a dielectric window. The coil parameters for the device were selected based on auxiliary experiments: coil diameter D = 12 mm, number of turns N = 6, width of the conductor W = 0.3 um, gap between turns G = 0.2 mm. The calculated inductance corresponding to these parameters is 337 nH. The coil and the rest of the electrical circuit, providing impedance matching with the RF source, were placed in a cylindrical sapphire case (see Figure 3). The coil was tightly glued to the flat end of the case. The thickness of the end wall was as thin as 0.3 mm so that the coil could provide a field strong enough to ignite and maintain the plasma at a moderate feeding power, i.e. without overheating the conductors. The sapphire case was hermetically sealed from the back by a stainless-steel flange equipped with an RF connector. The glue used in the assembly of the device is stable in vacuum conditions up to 500 K, the rest of the components placed in sapphire allow heating up to 400 K. To maintain a plasma in hydrogen at a pressure of 3-5 Pa, 1–2 W of RF power is sufficient, but a much higher power is required to ignite the plasma, at least 50 W applied for a few milliseconds. It is also worth mentioning that the constantly supplied power is limited by the thermal conductivity of the low-pressure hydrogen atmosphere and for the given device dimensions that does not exceed 6 W.

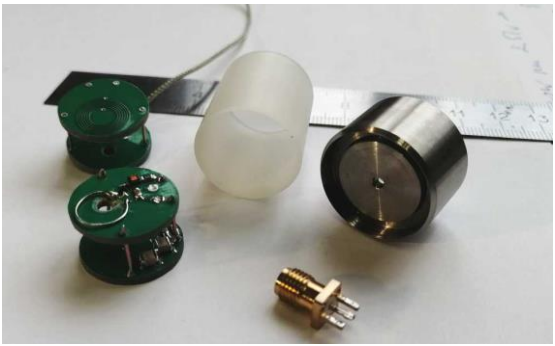
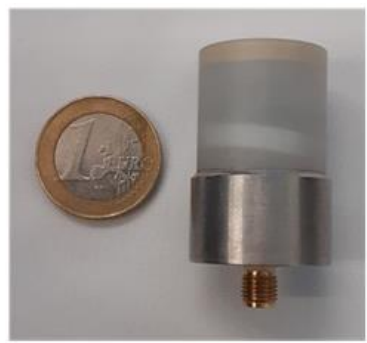

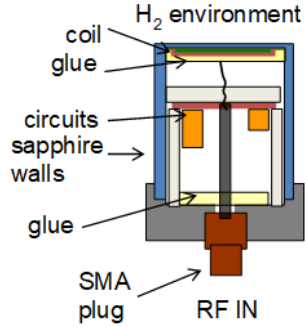

**Figure 3: Mini-ICP device assembly.**

The complete electrical circuit of the device consists of a master oscillator, a buffer amplifier, a power amplifier and a resonant circuit with the planar spiral coil (Figure 4). The master oscillator is based on the Colpitts circuit scheme with a transistor (see transistor Q1 in Figure 4). The varicap D1 is included in the master oscillator, which allows the oscillator frequency to be tuned in the 130-150 MHz range using the resistor R1. The buffer amplifier, assembled on transistors Q2 and Q3, serves to match the master oscillator with the power amplifier. The RA60H1317 module by Mitsubishi Electric was used as the power amplifier. This module provides up to 60 W of the output power in the 135-175 MHz frequency range. The output power can be adjusted by changing the voltage at the gates of the transistors of the module (pin 2). Since full power is only required when the plasma is ignited, the circuit uses a voltage regulator U2. Its output voltage is 3.7-4.1 V, which provides an amplifier output power of up to 10 W. When the S1 button is pressed, the output voltage for a short time (0.1 s) increases to 5.5 V, the output power rises to 60 W, which allows the plasma to be ignited. Then the output power drops to a value sufficient to maintain the plasma discharge.

The resonant circuit with a flat spiral coil L4 is mounted on a board made of low dielectric loss material. The resonant frequency of the circuit is about 130 MHz and is determined mainly by the inductance of the L4 coil and the capacitance of the capacitors C19 and C20. Capacitor C18 is required to match the circuit impedance at the resonant frequency to the impedance of the cable. Since the circuit has a high Q factor, it is necessary to control the fine tuning of the oscillator frequency to the resonant frequency of the circuit. For this, a coupling coil L5 is used, which has 1 turn and measures the voltage in the oscillatory circuit. The measured voltage is transmitted in the opposite direction along the coaxial cable and monitored at the U monitor output.

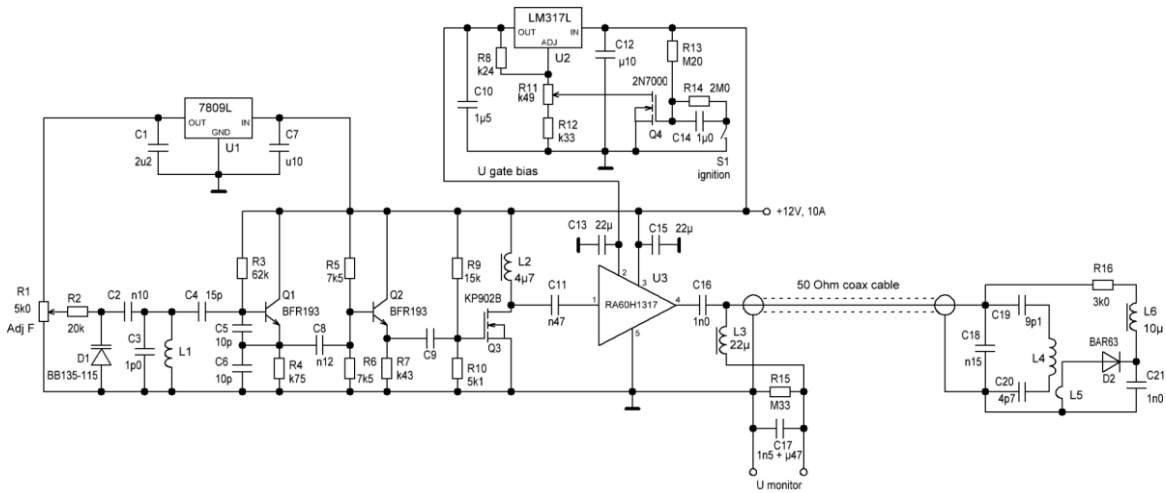

**Figure 4: Circuit diagram of the mini-ICP device.**

### III. PLASMA CHARACTERIZATION

The characterization of the plasma generated by the mini-ICP was carried out inside a vacuum vessel. A schematic representation of the test setup is given in Figure 5. As plasma metrology, we employed a Retarding Field Energy Analyzer (RFEA) from Impedans (Semion model) operating with a 3-grid button probe. The mini-

ICP was held vertically by a supporting structure hovering above the RFEA holder at an adjustable distance *D*.

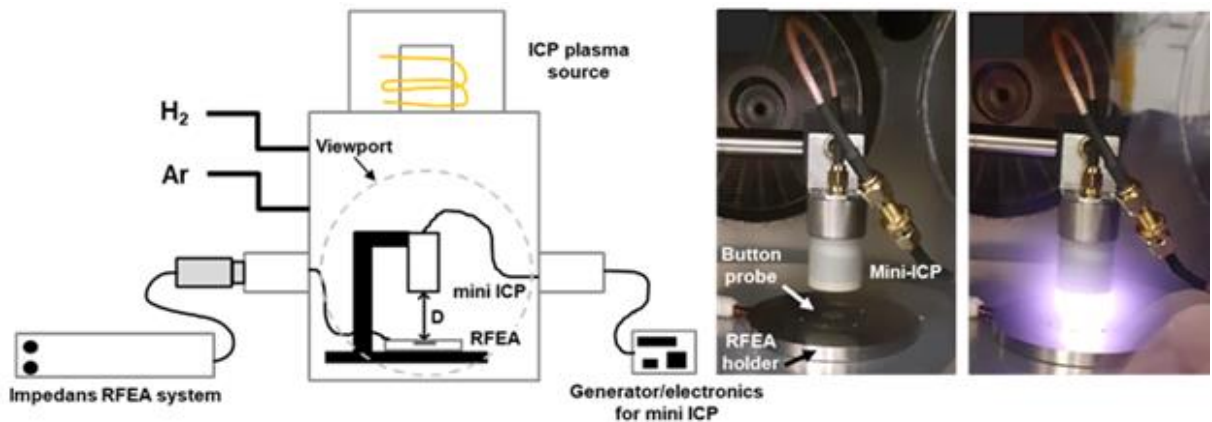

**Figure 5: Left: Schematic of the experimental setup used for plasma characterization. The employed vacuum vessel was equipped with feedthroughs for the RFEA & mini-ICP electronics as well as with an ICP plasma source located at the top (see main text for details). Middle and right: Photographs taken through the vessel viewport with the plasma on/off. Visible are the mini-ICP and the RFEA holder with its button probe installed in the middle.**

The vessel was equipped with hydrogen and argon gas supplies, as well as with an independent ICP plasma source positioned at the top (approx. 40 cm away from the mini-ICP). Measurements were taken at three different values of distance *D* (5, 12 and 50 mm) for a series of hydrogen pressures. For each set of conditions, pressure and distance, the source was operated at resonance (power estimated to be in the order of a few watts). In Figure 6 we plot fluxes versus pressure at the three different distances. For D=12 & 50 mm we observe a slight increase in flux with pressure followed by an apparent saturation towards high pressures (more clear for D=50 mm). For D=5 mm the flux seems rather insensitive to pressure within the tested range and it is not obvious to derive any trends. We note that there is no measurements for D=5 mm below 10 Pa as it was not possible to ignite the source; this is explained by the closeness of the RFEA grid resulting in loss of electrons to the surface before they are able to cause secondary ionizations.

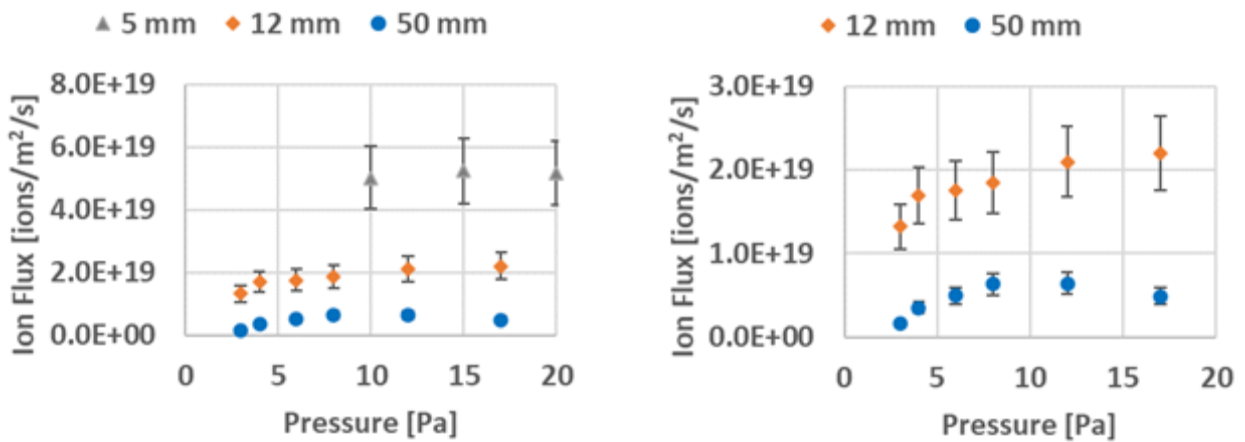

**Figure 6: Left: Flux versus pressure for D=5, 12, and 50 mm. A 20% error bar is shown to indicate the typical measurement reproducibility. Right: Zoom-in on the left-hand side data omitting the 5 mm points.**

In addition to flux measurements we have recorded the ion energy distribution function (IEDF) for a similar range of pressures versus distance. In Figure 7 we show a series of IEDF curves obtained at D=12 mm and pressures ranging from 3 to 17 Pa. Similar curves were recorded for D=5 mm and 50 mm (not shown here), and from these data we have produced the second plot in Figure 7. Two observations are clear: (i) for a given pressure, the ion energies increase the closer the source is to the RFEA. This is as expected since magnetic field is most confined and highest closest to the source coil, resulting in higher electron temperature; (ii) For any given distance, the peak energy decreases with pressure. This is be expected given that at higher pressures the number of collisions increases, hence lowering the electron temperature.

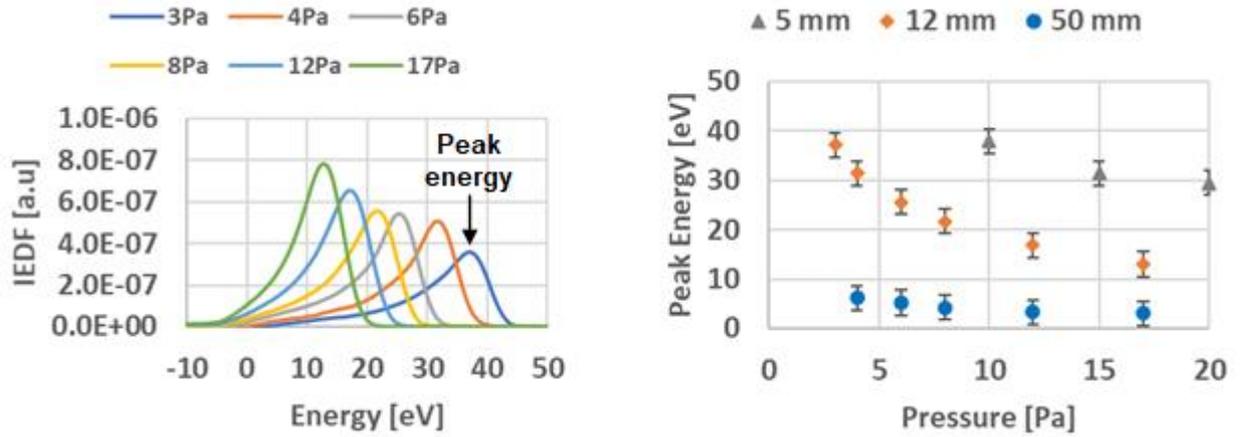

Figure 7: Left: Ion energy distribution functions (IEDF) as measured by the RFEA for different pressures and at D=12mm. For higher pressures a non-zero IEDF signal is observed at "negative ion energies". We are currently investigating the root cause of this, and we believe at this point that this is caused by spurious secondary electrons generated at the collector of the RFEA button probe upon ion impingement. Right: Peak energy versus pressure for D=5, 12 and 50mm. Peak energy is defined as indicated on (a) as the energy of the IEDF curve maximum.

As can be seen from Figure 6 the ion flux decreases with increasing gap, while Figure 7 shows the ion energy also decreases with increasing gap. However, if we include the additional constraint of keeping peak ion energies below 20 eV to prevent surface damage risks, the sweet spot appears to be at pressures of ~5-15 Pa and a gap distance of ~1-2 cm. The resulting ion flux and ion energy of ~$2 \cdot 10^{19}$ /m²s and ~20 eV respectively correspond to a carbon cleaning rate of ~2 nm/min, and a heat load of ~0.1 kW/m².

IV. PLASMA IGNITION

As stated above, it is hard to ignite the hydrogen plasma at low pressure and small gap, as shown in Figure 1 for several gap distances and pressures. This is due to the mean free path of the electrons becoming larger than the typical length scale of the test chamber (in this case the gap distance between coil window and RFEA probe surface), so the majority of electrons is lost to the walls without participating in secondary ionizations to sustain the discharge. The second ICP source in our vessel proved to be helpful to ignite the mini-ICP when no direct ignition assisted by the power boost was possible. This works by temporarily offering additional free electrons until the feedback mechanism of electron emission by accelerated ions incident upon the cathode is sufficiently established [14].

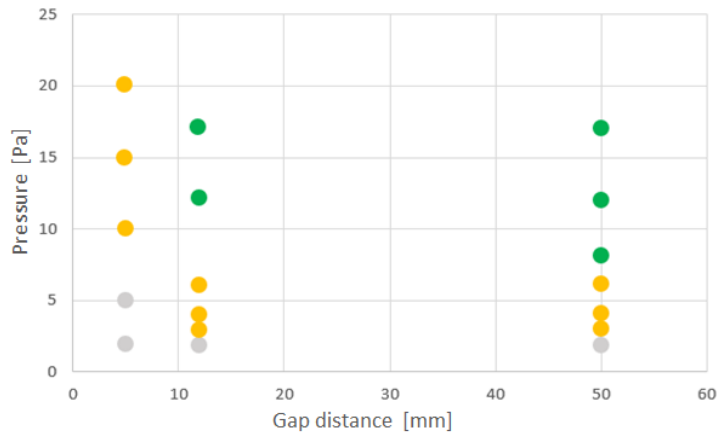

Figure 8: Summary of ignition behavior for different gap distances and pressures. Green: direct ignition ok (no need for secondary plasma). Orange: no ignition at the high-power pulse but with momentary help of the secondary plasma a discharge was sustained. Grey: no sustained plasma ignition at the high-power pulse (secondary plasma may lead to a discharge during high power pulse, but once the secondary plasma is off no discharge is sustained).

Besides a second plasma, ignition at lower pressure can be enhanced by several means, such as (i) changing window material to quartz instead of sapphire (improving in-coupling of power by lower permittivity of the window material), (ii) adding magnets to create helical trajectories for the electrons, thereby increasing the effective electron path length in the gas [15], (iii) matching the driving frequency to the electron collision frequency [13], (iv) photoelectric electron emission by UV-LED [16,17]. Other options such as piezo-induced spark or thermionic emission from a sharp needle are possible in theory, but are not considered due to concerns over particle generation.

Figure 9 shows an example of a modified mini-ICP with magnets integrated in the head behind the planar coil, and successful ignition at 4.5 Pa. The other ignition options were not tried yet for this specific application, but are considered well-proven in other application areas.

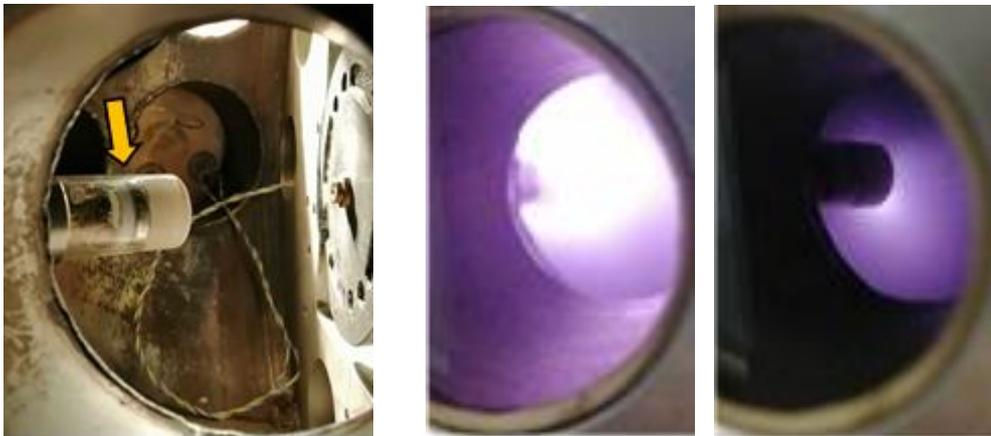

**Figure 9: Ignition at 4.5 Pa by magnets behind the coil and 60 W power boost (bright glow in middle picture), then gradual decrease to 2W at which the plasma is sustained (right picture).**

## V. Discussion and conclusion

The mini-ICP concept has been demonstrated to work well for pressures of 3 Pa and up, and gap distances of 1-5 cm. The resulting ion flux and energy of ~$2 \cdot 10^{19}$ /m$^2$s and ~20 eV correspond to a carbon cleaning rate of in order of 2 nm/min, and a heat load in order of 0.1 kW/m$^2$. This compares favorably against the HRG baseline, especially in cumulative heat load.

As expected, the ion density is relatively low compared to a full-size ICP plasma source; this is attributed to the relatively low impedance ratio of plasma to coil, and the lower volume-to-surface ratio. Still, the high in-coupling efficiency allows the device to run at an electrical power in the order of 1 W, so water-cooling is not needed even in (near) vacuum conditions.

To increase the cleaning rate, the ion density may be increased further by increasing the power and/or the driving frequency [18]. Such an increase in ion density will also keep the ion energy low to minimize the risk of sputtering [13].

Plasma ignition at low pressures and/or low gap distances requires assistance, but several proven ignition enhancements have been identified, and e.g. magnets have been proven to allow ignition below 5 Pa.

Besides cleaning, this kind of 'mini-ICP' plasma, with its small form factor and high in-coupling efficiency, may be applied to other fields such as de-charging wafers or reticles (or other substrates) and preventing charge build-up in dry (near) vacuum environments. Depending on the form factor requirements, the concept is quite versatile and even allows a truly 2D electronics design with lithographically formed inductors, resistors and capacitors on a 2D PCB (printed-circuit-board).

## VI. Acknowledgements

The authors wish to thank the ASML Research team for Scanner Plasma and Defectivity, and especially Ruud van der Horst, for fruitful discussions and general support.